\documentclass[amsmath,amssymb,aps,superscriptaddress,twocolumn]{revtex4-1}

\usepackage{amsmath,amssymb}
\usepackage{graphicx}
\usepackage[colorlinks=true,urlcolor=blue,breaklinks]{hyperref}
\usepackage{dcolumn}
\usepackage{hhline}
\usepackage{bm}
\usepackage{stmaryrd}
\usepackage[dvipsnames]{xcolor}
\usepackage[version=4]{mhchem}
\usepackage{siunitx}
\usepackage{array}
\newcolumntype{M}[1]{>{\centering\arraybackslash}m{#1}}
\usepackage{paralist}
\usepackage{enumitem}
\usepackage{natbib}

\def\sio2{{\texorpdfstring{SiO$_2$}{}}}

\DeclareMathAlphabet\mathbfcal{OMS}{cmsy}{b}{n}
\def\d{{\rm d}}
\def\g{{g}}
\def\etal{\emph{et al}}

\begin{document}

\title{Yield stress aging in attractive colloidal suspensions}

\author{Francesco Bonacci}
\affiliation{PMMH, CNRS, ESPCI Paris, Université PSL, Sorbonne Université, Université de Paris, F-75005, Paris, France}
\author{Xavier Chateau \thanks{xavier.chateau@enpc.fr}}
\affiliation{Navier, Ecole des Ponts, Univ Gustave Eiffel, CNRS, Marne-la-Vallée, France}
\author{Eric M. Furst}
\affiliation{Department of Chemical and Biomolecular Engineering, University of Delaware, 150 Academy Street, Newark, Delaware, 19716, USA}
\author{Julie Goyon}
\affiliation{Navier, Ecole des Ponts, Univ Gustave Eiffel, CNRS, Marne-la-Vallée, France}
\author{Ana\"el Lema\^{\i}tre \thanks{anael.lemaitre@enpc.fr}}
\affiliation{Navier, Ecole des Ponts, Univ Gustave Eiffel, CNRS, Marne-la-Vallée, France}

\date{\today}

\begin{abstract}
We investigate the origin of yield stress aging in semi-dense, saline, and turbid suspensions in which structural evolution is rapidly arrested by the formation of thermally irreversible roll-resisting interparticle contacts. By performing optical tweezer (OT) three-point bending tests on particle rods, we show that these contacts yield by overcoming a rolling threshold, the critical bending moment of which grows logarithmically with time. We demonstrate that this time-dependent contact-scale rolling threshold controls the suspension yield stress and its aging kinetics. We identify a simple constitutive relation between the contact-scale flexural rigidity and rolling threshold, which transfers to macroscopic scales. This leads us to establishing a constitutive relation between macroscopic shear modulus and yield stress that is generic for an array of colloidal systems.
\end{abstract}
\maketitle

Our understanding of aging in attractive colloidal suspensions has been guided for decades by studies of transparent and sterically stabilized experimental models~\cite{KilfoilPashkovskiMastersWeitz2003,DibbleKoganSolomon2006} in which, by construction, van der Waals forces are absent and contact formation is excluded. In such systems, aging results from a slow (glass-like) evolution of the microstructure~\cite{AbouBonnMeunier2001,CipellettiRamos2005}. 

But most colloidal suspensions in the environment, industry, or in civil engineering, are saline and turbid. Turbidity signals the existence of an index contrast between particles and suspending fluid, i.e. of \emph{attractive} van der Waals forces; salinity introduces ions that screen particle charges, thus weakening the repulsive Coulombic forces. These two properties hence conspire to facilitate the formation of solid-solid interparticle contacts.

It was pointed out very recently that moderate levels of ionic strength and index contrast suffice to bias the balance between Coulombic repulsion and van der Waals attraction to the point that no repulsive barrier limits the formation of adhesive, roll-resisting, and thermally stable interparticle contacts, with the consequence that, at intermediate ($30\%$ to $40\%$) packing fractions, the microstructure freezes within seconds of flow arrest~\cite{BonacciChateauFurstFusierGoyonLemaitre2020}. But these systems do present mechanical aging on timescales up to hours~\cite{CoussotTabuteauChateauTocquerOvarlez2006,RousselOvarlezGarraultBrumaud2012,OvarlezCoussot2007,FusierGoyonChateauToussaint2018}, much beyond structural arrest. For a broad class of real-life suspensions, mechanical aging is hence non-structural, and governed by contact scale physical processes.

Identifying these processes is a major challenge, requiring joint advances in experiments (spotting and characterizing the relevant contact scale processes) and theory (rationalizing the underlying physical mechanisms; scaling them up from contact to macroscopic properties). Macroscopic shear modulus ($G'$) aging can be related to the growth of contact-scale flexural rigidity~\cite{BonacciChateauFurstFusierGoyonLemaitre2020}. But, it leaves wide open the most crucial issue in most practical situations: the mechanism of macroscopic yield stress $\sigma_y$ aging.

Most yield stress models for suspensions~\cite{ScalesJohnsonHealyKapur1998,FlattBowen2006} indeed assume interactions to be centro-symmetric. They hence do not offer any insight as to how roll-resisting contacts determine macroscopic modulus and yield stress aging. It was shown experimentally that interparticle flexural contacts yield by overcoming a rolling threshold~\cite{PantinaFurst2008}, but no experimental evidence exists for microscopic yield aging at the scale of adhesive contacts between colloidal particles. And no direct, quantitative link has ever been established between contact and macroscopic yielding in a given attractive colloidal suspension.

Here, we address these interrelated issues by investigating yielding at both the macroscopic and contact scales in aqueous suspensions of St\"ober silica particle flocculated by addition of \ce{CaCl2} at moderate ionic strengths. By performing three-point bending tests with OTs on particle rods, we show that contacts yield when reaching a rolling threshold. We then demonstrate that the associated critical bending moment $M_y$ ages and---by comparison with macroscopic rheometry data---that this contact-scale process controls macroscopic yield stress aging with $\sigma_y\propto M_y$. This yields a constitutive relation between the macroscopic shear modulus and yield stress $\sigma_y(t_w)\propto \sqrt{G'(t_w)}/a$, with $t_w$ the age and $a$ the particle radius, the prefactor being a material-dependent constant of unit N$^{1/2}$. One consequence of our finding is that it is possible to track the growth of the yield stress by monitoring the shear modulus, an observation with far-reaching consequences for real-life situations.\\

\begin{figure}[!t]
  \centering
  \includegraphics[scale=0.5]{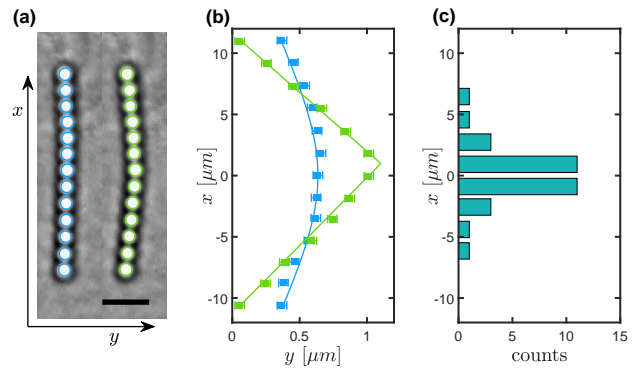}
  \caption{(a) Two snapshots of a 13 silica particle rod, just before and just after a yielding event. The colored circles show the particle positions as reconstructed from subpixel image analysis (radii are reduced for better legibility). Scale bar $= \SI{5}{\micro\meter}$. (b) The reconstructed particle positions in the $(x,y)$ plane, superposed, after magnification along the $y$ axis. Pre-yielding positions (blue) agree with the Euler-Bernoulli equation (line); post-yielding ones (green) form two straight segments connected at a finite angle. (c) Distribution of first yield  event locations ($x^\star$): $\sim70\%$ occur near the rod center, where the bending moment is maximum.
  }
  \label{fig:bending}
\end{figure}

Interparticle contacts are probed using Pantina and Furst's method~\cite{PantinaFurst2005,PantinaFurst2006}, which consists in a three point bending test on a rod comprising an odd number of particles (see Fig.~\ref{fig:bending}a). During a test, two fixed traps hold the rod extremities; a third one grabs the central particle and is then translated perpendicularly to the rod, at a velocity slow enough to avoid hydrodynamic drag effects. Sample preparation and measurement protocol follow the Method section of Ref.~\cite{BonacciChateauFurstFusierGoyonLemaitre2020} except otherwise stated below.

Here, we perform these tests using particles of diameter $2a=\SI{1.9}{\micro\meter}$. We find contacts to be thermally stable at all the considered ionic strengths---our rods keep their integrity for hours when held by the two end traps~\cite{BonacciChateauFurstFusierGoyonLemaitre2020}. Also, we find that two particles brought into contact cannot be pulled apart using optical traps---we tried, but failed, while developing forces up to about 15pN, i.e. $\sim3$ times larger than those we will exert here. Consistently, the Derjaguin-Muller-Toporov (DMT)~\cite{DerjaguinMullerToporov1975} estimate places the contact pull-off force $\sim\pi a W \simeq\SI{24}{nN}$~\footnote{The work of adhesion estimates as $W \simeq \frac{A_H}{12 \pi D_0^2}\approx\SI{8}{mJ/m^2}$ with $A_H=8.3\times 10^{-21}\SI{}{J}$, the non-retarded Hamaker constant of silica in water and estimating the surface-surface separation at contact to be $D_0 = \SI{0.165}{nm}$~\cite{Israelachvili2011}.}, much beyond the range accessible to OTs.

Force and deflection measurements rest on image analysis with subpixel resolution~\cite{CrockerGrier1996}. Before each test, the average trap stiffness $k$ is measured by monitoring the thermal fluctuations of particles held in our three traps. Once the rod formed, before loading, the average positions of both end particles are measured to identify the end trap locations. The line connecting these two points defines the $x$ axis. Loading is performed along the transverse and horizontal axis, denoted $y$. The force exerted on the central particle is obtained, in essence, as $f =k (\Delta y_1+\Delta y_N)$, with $\Delta y_1$ and $\Delta y_N$ the y-displacements of both end particles from the end trap locations. Finally, the rod deflection $\delta$ is measured as the difference between the $y$ coordinates of the center and end particles. In practice, rods are not strictly linear: the associated misalignments are corrected by analyzing the 3D rod structures~\cite{BonacciChateauFurstFusierGoyonLemaitre2020}.\\

\begin{figure}[!t]
  \centering
  \includegraphics[scale=0.5]{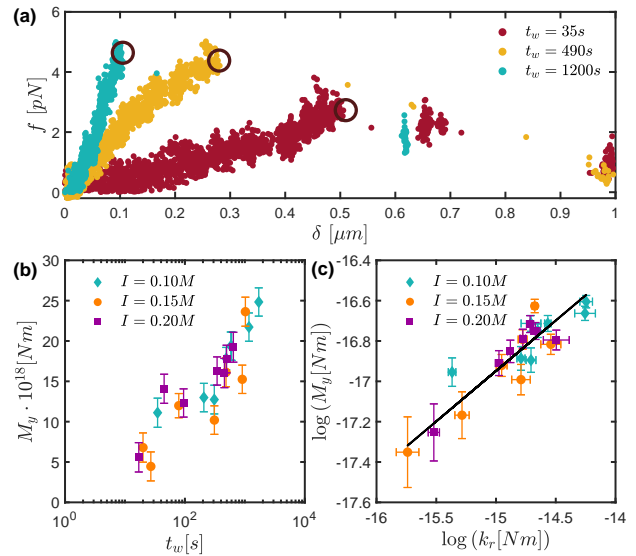}
  \caption{(a) Typical bending force ($f$) vs deflection ($\delta$) curves after three aging times. Circles mark the yield points $(f_y,\delta_y)$. (b) Log-lin plot of the critical bending moment ($M_y$) versus aging time ($t_w$). 
    Error bars are deduced from measurement uncertainties in $f_y$ and $x^\star$.
    (c) Critical moment ($M_y$) versus flexural stiffness ($k_r$) for all ages and ionic strengths.}
\label{fig:contact}
\end{figure}

At chosen aging times ($t_w$, as counted starting from the formation of the last bond in the assembly process) we perform flexural tests and monitor force vs deflection as illustrated in Fig.~\ref{fig:contact}a.

At small deflections, rods respond elastically with $f$ increasing essentially linearly with $\delta$. Meanwhile, the rod deformation is well-described by the Euler-Bernoulli beam equation (solid blue line in Fig.~\ref{fig:bending}b), which entails that contacts support finite torques, i.e. resist rolling~\cite{BonacciChateauFurstFusierGoyonLemaitre2020}. The associated effective bending rigidity is $k_0=8(N-1)^3f/\delta$, with $N$ the number of beads. Modelling the rod as a series of beads connected by roll-resisting contacts, yields the contact scale flexural stiffness $k_r=k_0a^2/\Gamma$ with $\Gamma\simeq96$ a slightly $N$-dependent parameter~\cite{BeckerBriesen2008}. These stiffnesses grow quasi-logarithmically with time~\cite{BonacciChateauFurstFusierGoyonLemaitre2020}, which attests to the existence of contact-scale aging dynamics.

With the increasing deflection, yielding eventually occurs, quite abruptly, without any evidence of incipient plastic activity, at a yield point $(f_y,\delta_y)$.
Immediately afterwards (Fig.~\ref{fig:bending}a,b, green data), the rods systematically display a triangular shape, which evidences that a single contact (the apex, with abscissa $x^\star$), has rolled. The rods do not break open after yielding; they can still be held by traps and be reloaded, repetitively, through other similar yield events [Fig.~\ref{fig:contact}a].\\

To shed light on the yielding mechanism, we perform a large number of tests and report the $x^\star$ distribution in Fig.~\ref{fig:bending}c. If yielding resulted from frictional sliding, the distribution of $x^\star$ would be uniform, because the shear force along a bent rod is. In contrast, we find that yield events occur overwhelmingly at $x^{\star}=\pm a$, i.e. in the contacts formed by the central particle, which is where the local moment $M_y = ({f_y}/{2})\left[ {L}/{2} -  \vert x^\star \vert \right]$ is maximal. These data hence unambiguously demonstrate that yielding results from the crossing of a rolling threshold~\cite{PantinaFurst2008}. 

Accordingly, the rare yield events occurring away from the center particle must be attributed to experimental artifacts, such as contact-scale defects or poorly formed rods. Therefore, in order to carry out a quantitative, time-resolved analysis of roll yielding, we only retain the $x^{\star}=\pm a$ events, and report $M_y$ vs $t_w$ in Fig.~\ref{fig:contact}b. Remarkably, $M_y$ grows roughly logarithmically at late times, like the flexural stiffness $k_r$ does~\cite{BonacciChateauFurstFusierGoyonLemaitre2020}. It is also essentially $I$-independent in the studied ionic strength range, over which the charge carried by our particles is constant~\cite{FusierGoyonChateauToussaint2018,BonacciChateauFurstFusierGoyonLemaitre2020}.

It was never previously reported that the contact scale rolling threshold grows logarithmically in time. This observation hence constitutes a key experimental finding about adhesive colloidal suspensions.

Plotting $M_y$ vs $k_r$ as parametrized by $t_w$ [Fig.~\ref{fig:contact}c], we find that these two contact properties quite nicely obey the relation (solid line)
\begin{equation}
\label{eq:My}
M_y = \g\,k_r^\alpha
\end{equation}
with $\alpha = 1/2$ and $\g=(3.56\pm0.3)\cdot10^{-10} \SI{}{N^{1/2}m^{1/2}}$. This relation implies that the critical bending angle $\theta_y=M_y/k_r=\g/\sqrt{k_r}$ decreases with $t_w$, i.e. that \emph{contacts become increasingly brittle as they age}. This is directly visible in Fig.~\ref{fig:contact}a as $\delta_y$ clearly decreases with $t_w$. It points to a contact yield mechanism (contact line depinning) akin to fragile rupture.

In this OT study, we focused on a single particle size, due to the difficulties in accumulating data points. While $g$ is a priori $a$-dependent, the proximity of the measured exponent $\alpha$ to $1/2$ suggest it might be prescribed by simple physical principles, in which case the scaling $M_y(t_w)=g\times[k_r(t_w)]^{1/2}$ would constitute a generic property of the contacts between microspheres. We explore this issue in the rest of the paper.\\

Rolling resistance signals that, at any aging time, when a contact responds elastically, its contour remains pinned. As flexural rigidity results from elastic strains inside the particles and is determined by the contact geometry, its aging unambiguously demonstrates the growth of the contact radius $a_c$. Following Furst~\etal~\cite{PantinaFurst2005}, we estimate $k_0 = 12\pi E a_c^4/a^3$~\cite{LandauLifshitz1986}, the bending stiffness of a rod of diameter $a_c$ and Young's modulus $E$, which yields:
\begin{equation}\label{eq:kr}
  k_r=\frac{12\pi E a_c^4}{a\Gamma}
\end{equation}
This $a_c^4$ scaling is supported by shear modulus aging data for a range of particle sizes~\cite{BonacciChateauFurstFusierGoyonLemaitre2020}. It arises because flexion introduces a linear stress $\sigma\propto\theta\,y$ throughout the contact area $A_c$ ($y$ being the bending direction), so that integrating the associated torque yields $\int_{A_c}\d y \d z\,y\sigma\sim a_c^4\,\theta$.

The growth of $a_c$ points to a type of sintering process (e.g. the progressive formation of siloxane bridges~\cite{Gauthier-ManuelGuyonRouxGitsLefaucheux1987,VigilXuSteinbergIsraelachvili1994,LiuSzlufarska2012}), which increases the overall cohesion energy inside the contact area. Thus, here, adhesion results from two distinct types of interactions: van der Waals attraction and intra-contact bonds formed by sintering. The former is time-independent and invariant under particle rotations: it brings about contact formation and adhesion, yet \emph{does not introduce rolling resistance}, which explains that contacts do not display measurable flexural rigidity immediately after they form~\cite{BonacciChateauFurstFusierGoyonLemaitre2020}. Intra-contact bonding is time-dependent and begets rolling resistance.


We construct a schematic description of such a contact, in the spirit of contact theories~\cite{Barthel2008},
all of which relate the contact diameter to the adhesion energy $W$ via:
\begin{equation}\label{eq:ac}
a_c(t_w)= A\left(\frac{3\pi\,a^2\,W(t_w)}{8E^*} \right)^{1/3}
\end{equation}
with $E^*=E/(1-\nu^2)/2$ the reduced modulus, $E=\SI{30}{GPa}$~\cite{PaulRomeisTomasPeukert2014}, $\nu=0.17$ the Poisson's ratio~\footnote{Even large changes in $\nu$ have little effect on $a_c$ as $1/(1-\nu^2)^{1/3}$ varies from 1 to 1.1 when $\nu$ varies from 0 to 0.5. We can thus safely use an average value for fused silica's Poisson ratio~\cite{PaulRomeisTomasPeukert2014}.}. For our aging contacts, $W$ should be interpreted as a $t_w$-dependent effective adhesion energy, which integrates all adhesive forces inside the contact. The precise value of $A$, $=1$ resp. $3^{1/3}\simeq1.44$ in DMT and Johnson–Kendall–Roberts (JKR)~\cite{JohnsonKendallRoberts1971} theories, is irrelevant to our analysis.

Next, since adhesion arises from atomic-scale forces developed by sintering, the onset of rolling is expected to be determined by a fragile rupture criterion.
Namely~\cite{KrijtDominikTielens2014}, roll-yielding should occur when the strain energy release rate during rolling, $\Delta G$, equals the adhesion hysteresis $\Delta W$, i.e. the difference between the surface creation and opening energies at the leading and trailing edges (resp.). After calculating $\Delta G$ as a function of the bending level for a JKR contact, Krijt~\etal~\cite{KrijtDominikTielens2014} thus obtain the following yielding criterion
\begin{equation}\label{eq:thetay}
  \theta_y(t_w)=\frac{a_c(t_w)\,\Delta W}{6\,a\,W(t_w)}
\end{equation}
where we explicitly write out all the $t_w$-dependencies. In Krijt~\etal's calculation, $W$ appears via the global condition of zero total force: it corresponds to the $W(t_w)$ of Eq.~(\ref{eq:ac}), which integrates all age-dependent adhesive contributions throughout the contact. In contrast, we expect the adhesion hysteresis $\Delta W$ to be $t_w$-independent because (i) the closing energy obviously is; (ii) the opening energy too because, at the onset of yielding, opening occurs at the rim of the growing contact where sintering has not taken place yet.

In our model, the time-dependent state of a contact reduces to the contact radius $a_c(t_w)$, which relates to the effective adhesion energy via $W(t_w)\sim a_c^3(t_w)$ [Eq.~(\ref{eq:ac})]. Equation~(\ref{eq:thetay}) then yields $\theta_y(t_w)\sim1/a_c^2(t_w)$, all coefficients being constant. Since $k_r(t_w)\sim a_c^4(t_w)$ [Eq.~(\ref{eq:kr})], we obtain $\theta_y(t_w)\sim1/\sqrt{k_r(t_w)}$, or $M_y(t_w)\sim\sqrt{k_r(t_w)}$, i.e. Eq.~(\ref{eq:My}) with the prefactor:
\begin{equation}\label{eq:g}
  g=\frac{(3\pi)^{3/2}A^3(1-\nu^2)\Delta W}{12\sqrt{\Gamma E}}\sqrt{a}
\end{equation}
Using the JKR expression $A^3=3$, the measured $g\simeq 3.56\cdot10^{-10} \SI{}{N^{1/2}m^{1/2}}$ corresponds to $\Delta W\simeq\SI{77}{mJ/m^2}$, a quite reasonable value~\footnote{For a typical surface energy of silica in water $W_0\simeq\SI{100}{mJ/m^2}$~\cite{Iler1979}, this yields $\Delta W/W_0\simeq0.77$, while Krijt et al~\cite{KrijtDominikTielens2014} report $\Delta W/W_0$ values ranging between 0.5 and 1 for silica.}, which provides compelling support to our argument.\\

For a fixed microstructure, we expect the shear modulus~\cite{BonacciChateauFurstFusierGoyonLemaitre2020} and yield stress to be proportional to the flexural stiffness $k_r$ and critical moment $M_y$ respectively. For dimensional reasons, these macroscopic properties then relate to their microscopic counterparts via:
\begin{align}
\label{eq:elasticmodulus}
G'(a,\phi, t_w) &= \frac{S(\phi)}{a^3}  \times k_r(a, t_w)\\
\label{eq:sigmay}
  \sigma_y(a,\phi, t_w)&=  \frac{Q(\phi)}{a^3} \times M_y(a, t_w)
\end{align}
where $S$ and $Q$ characterize the microstructure, and are hence independent of time, particle size, and ionic strength.

Equation~(\ref{eq:sigmay}) offers us an opportunity to test our theoretical analysis of the contact problem, which predicts the non-trivial scaling $M_y(a,t_w)\sim W^{2/3}(t_w)\,a^{4/3}$. This cannot be done directly using OTs due to the limited range of accessible particle sizes. But, together with Eq.~(\ref{eq:sigmay}), it predicts that, for two different radii $a$ and $a^\star$: 
\begin{equation}\label{eq:test}
\sigma_y(a,\phi,t_w)=\frac{Q(\phi)}{a^3}\,\left(\frac{a}{a^\star}\right)^{4/3}\,M_y(a^\star,t_w)
\end{equation}
which expresses the macroscopic threshold of suspensions of arbitrary $a$ as a function of $M_y(a^\star,t_w)$.

\begin{figure}
  \centering
  \includegraphics[scale=0.5]{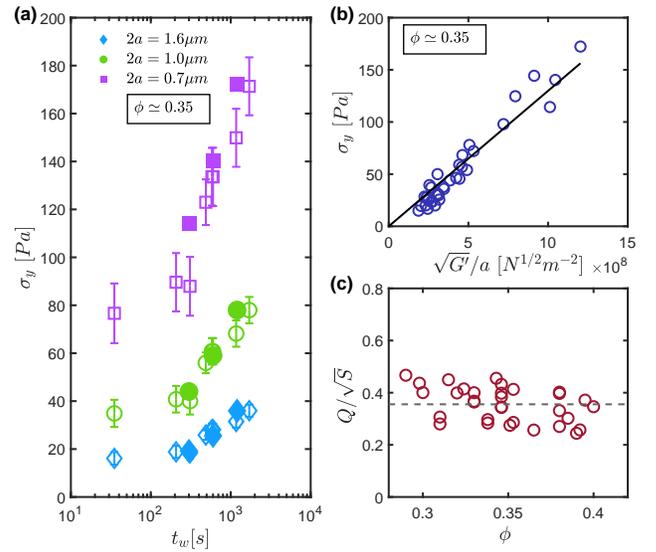}
  \caption{(a) Macroscopic yield stress (filled symbols) of $\phi\simeq0.35$ suspensions as measured from rheology stress sweep tests. Open symbols (in corresponding colors) are the predictions obtained from Eq.~(\ref{eq:test}) using the single value $Q\simeq 0.9$ for all three series of points.
    (b) Test of the macroscopic constitutive relation [Eq.~(\ref{eq:macro})]: $\sigma_y$ vs $\sqrt{G'}/a$ for all $\phi\simeq0.35$ suspensions, with $2a$ ranging from 0.7 to $\SI{1.6}{\micro\meter}$ and $I$ from 0.1 to $\SI{0.2}{M}$. The solid line is not a fit, but the prediction of Eq.~(\ref{eq:macro}) using the $\phi$-average value of $Q/\sqrt{S}\sim0.355$ from panel~(c) and the fitted value of $g(a^{\star})/\sqrt{a^{\star}}$.
    (c) $Q/\sqrt{S}$ as a function of $\phi$, obtained from independent fits of $Q$ and $S$ by matching microscopic and macroscopic data.}
\label{fig:sigmay}
\end{figure}

We have systematically tested this relation against rheometry data~\cite{FusierGoyonChateauToussaint2018}. Like $M_y$, $\sigma_y$ does not depend on $I$, but strongly grows with $\phi$ over the studied range. 
A typical test is presented in Fig.~\ref{fig:sigmay}a, where filled symbols represent $\sigma_y(t_w)$ for $\phi\simeq0.35$ suspensions, at aging times $t_w=300$, 600, \SI{1200}{s}, and for $2a=0.7$, 1.0, and \SI{1.6}{\micro m}. As seen, $\sigma_y$ increases with $t_w$ and decreases with $a$. Testing Eq.~(\ref{eq:test}) implies being able to reconstruct both the $t_w$ and $a$-dependence of these data points using our $M_y(a^\star,t_w)$ OT data, with $2a^\star=\SI{1.9}{\micro m}$ [Fig.~\ref{fig:contact}], and a single free parameter $Q$.
This highly constrained fit yields the three series of open symbols: it works remarkably well, which brings decisive support to all of our analysis.\\

A major outcome of our observations and analysis is that aging contacts obey the microscopic constitutive relation $M_y=g\sqrt{k_r}$ [Eq.~(\ref{eq:My})].
Even more importantly, combining Eqs.~(\ref{eq:My}),~(\ref{eq:g}),~(\ref{eq:elasticmodulus}) and~(\ref{eq:sigmay}), we now predict that
\begin{equation}\label{eq:macro}
  \sigma_y(a,\phi, t_w)=C(\phi)\,\frac{\sqrt{G'(a,\phi, t_w)}}{a}
\end{equation}
which is a constitutive relation between the macroscopic shear modulus and yield stress of an aging suspension. Here, $C(\phi)=(g(a)/\sqrt{a})\,Q(\phi)/\sqrt{S(\phi)}$ is a function of $\phi$ only since $g/\sqrt{a}$ only depends on physical properties of the particles [Eq.~(\ref{eq:g})]. This relation is tested in Fig.~\ref{fig:sigmay}b where we plot $\sigma_y$ vs $\sqrt{G'}/a$ for $\phi\simeq0.35$ and an otherwise broad range of conditions ($2a$ from 0.7 to $\SI{1.6}{\micro\meter}$, $I$ from 0.1 to $\SI{0.2}{M}$). The agreement is remarkable.

For various values of $\phi$ over the range (from $\phi\simeq0.3$ to 0.4) over which our suspensions are stable and present a measurable yield stress,
we have obtained $Q$ and $S$ independently: the former, by matching Eq.~(\ref{eq:test}) as in Fig.~\ref{fig:sigmay}; the latter via a similar analysis of $G'$ and $k_r$ data~\cite{BonacciChateauFurstFusierGoyonLemaitre2020}. The resulting values of $Q(\phi)/\sqrt{S(\phi)}$, presented in Fig.~\ref{fig:sigmay}c, do not show any systematic $\phi$-dependence.
This is quite meaningful because, although the accessible range of packing fractions is arguably limited, the parameters $S(\phi)$ and $Q(\phi)$ vary separately by significant factors: $S$ from 1.9 to 47 (a factor of $\simeq25$), and $Q$ from 0.5 to 2.5 (a factor of 5). 

What could be the cause of the $\phi$-independence of $Q/\sqrt{S}$? Observe that, despite experimental difficulties, our $k_r(t_w)$ data~\cite{BonacciChateauFurstFusierGoyonLemaitre2020} show quite moderate sample-to-sample fluctuations. Hence, within a given suspension at rest, the flexural modulus grows essentially at the same rate in all contacts. Thus, although contacts have slightly different times of formation, their age difference rapidly becomes negligible with the increasing ti/Users/xavier/Downloads/Yield stress aging in adhesive colloidal suspensions/biblio.bibme. Therefore, we can safely assume that, within a suspension at rest, beyond a short transient of order seconds, all contacts present the same age-dependent $k_r(t_w)$.

Normal stiffnesses being considered essentially infinite, the homogeneity of $k_r(t_w)$ guarantees that, when a suspension is elastically loaded, the resulting microscopic non-affine motions are time-independent.
Thus, at macroscopic strain $\gamma$ within the elastic regime, for an arbitrary contact $ij$, the flexion angle $\theta_{ij}=A_{ij}\gamma$, with $A_{ij}$ time-independent.
In a system of volume $V$, the energy density $E^{\rm el}=\frac{1}{2V}\sum_{ij} k_r A_{ij}^2\gamma^2$ where the sums run over all contacts. The elastic modulus
$G'=(1/\gamma)\d E^{\rm el}/\d\gamma=\frac{N_c}{V}\langle A_{ij}^2\rangle\,k_r$ with $N_c$ the number of contacts and $\langle\cdot\rangle$ the spatial (or ensemble) average.
Since we restrict our attention to a limited range of packing fractions away from jamming, we may write $N_c/V\simeq\rho/a^3$, with $\rho$ a constant of order a few units. The effect of packing fraction, hence, is entirely contained in the geometric factor $\langle A_{ij}^2\rangle$, and we recover Eq.~(\ref{eq:elasticmodulus}) with $S=\rho\langle A_{ij}^2\rangle$.

Now, let us state an additional evidence: suspensions are ductile. Yet, we just found that their yielding is determined by a brittle process: the depinning of contact lines, with contacts being increasingly fragile with age. Clearly, macroscopic yielding cannot result from that of a single or even a few contacts. It should instead occur when a measurable fraction of contacts are brought past their threshold, thus precipitating a drop in the elastic modulus. Under such yielding conditions, the rescaled moment $\sqrt{\langle \theta_{ij}^2\rangle}/\theta_y(t_w)=\kappa$ should achieve a constant value. Therefore, the yield strain $\gamma_y=\kappa\theta_y/\sqrt{\langle A_{ij}^2\rangle}$ and $\sigma_y=G'\gamma_y=(1/a^3)\sqrt{\rho\,S}\kappa M_y$. Not only do we now recover Eq.~(\ref{eq:sigmay}), but we also predict that $Q/\sqrt{S}=\kappa\sqrt{\rho}$ is a constant, which is exactly what our data supports.
\\

In summary, we have brought compelling evidence that adhesive colloidal suspensions yield by a mechanism that depends on interparticle contacts reaching an age-dependent rolling threshold associated with the depinning of the contact line, a brittle rupture mechanism. We have identified a microscopic constitutive relation [Eq.~(\ref{eq:My})] relating this threshold to the flexural stiffness and were able to explain its origin within a model which, although schematic, is fully consistent with both our previous interpretation of the origin of flexural rigidity~\cite{BonacciChateauFurstFusierGoyonLemaitre2020} and an existing estimate of the rolling threshold~\cite{KrijtDominikTielens2014}. This model, moreover, yields a prediction about the particle size dependence of the macroscopic yield stress, which we successfully tested.

This led us to identify a macroscopic constitutive relation between the yield stress and shear modulus [Eq.~(\ref{eq:macro}), with $C$ a constant], which constitutes a major outcome. It opens two perspectives of considerable practical interest: identifying non-destructive probes of the age-dependent yield stress of suspensions, the most important property in many situations; or controlling the yield stress (jointly with the modulus) by altering the surface chemistry of particles.

\bibliographystyle{apsrev4-1}
\bibliography{biblio}

\end{document}